\documentclass[12pt,usenatbib]{mn2e}
\usepackage{graphicx}
\usepackage{epsf}
\def\lsim{\mathrel{\lower0.6ex\hbox{$\buildrel {\textstyle <}
 \over {\scriptstyle \sim}$}}}
\def\gsim{\mathrel{\lower0.6ex\hbox{$\buildrel {\textstyle >}
 \over {\scriptstyle \sim}$}}}

\begin{document}

\title[Radial distribution of subhaloes and satellite galaxies]
      {Constrained simulations of the Local Group: on the radial distribution
      of substructures}
\author[Libeskind et al.] 
{Noam I Libeskind$^1$, Gustavo Yepes$^2$, Alexander Knebe$^2$, Stefan
      Gottl\"ober$^1$,  \newauthor Yehuda Hoffman$^3$, Steffen R Knollmann$^2$
  \\
  $^1$Astrophysikalisches Institut Potsdam, An der Sternwarte 16, D-14482 Potsdam, Germany\\
  $^2$Grupo de Astrof\'\i sica, Departamento de Fisica Teorica, Modulo C-XI, Universidad Aut\'onoma de Madrid, Cantoblanco E-280049, Spain\\
  $^3$Racah Institute of Physics, The Hebrew University of Jerusalem, Givat Ram, Israel
  }
%\date{Accepted 1988 December 15. Received 1988 December 14; in original form 1988 October 11}

%\pagerange{\pageref{firstpage}--\pageref{lastpage}} \pubyear{2002}

\maketitle \begin{abstract} 
We examine the properties of satellites found in high resolution
simulations of the local group. We use constrained simulations designed
to reproduce the main
dynamical features that characterize the local neighborhood, i.e. within tens of Mpc 
around the Local Group (LG). Specifically, a LG-like object is found located
within the 'correct' dynamical environment and consisting of three main objects which are associated with the Milky Way, M31 and M33. By running two simulations of this LG from identical initial
conditions 
- one with and one without baryons modeled hydrodynamically - we can quantify
the effect of 
gas physics on the $z=0$ population of subhaloes in an environment similar to
our own. We find that above a certain mass cut,
$M_{\rm sub} > 2\times10^{8}h^{-1} M_{\odot}$ subhaloes in
  hydrodynamic simulations are more radially concentrated than those in
  simulations with out gas. This is caused by the
  collapse of baryons into stars that typically sit in the central regions of
  subhaloes, making them denser. The increased central density of such a
  subhalo, results in less mass loss due to tidal stripping than the same
  subhalo simulated with only dark matter. The increased mass in hydrodynamic
  subhaloes with respect to dark matter ones, causes dynamical friction to be
  more effective, dragging the subhalo towards the centre of the
  host. This results in these subhaloes being effectively more radially
  concentrated then their dark matter counterparts.  

\end{abstract}
%\keywords{galaxy formation: general --- Black Holes, galaxy formation}

\section{Introduction}
\label{introduction} 
In the current paradigm of structure formation, large objects such as galaxies
or clusters, are believed to form
hierarchically, or through a ``bottom-up'' \citep{1978MNRAS.183..341W} process
of merging. Over dense peaks in the primordial density field collapse at high
redshift due to gravitational instability and then merge with each other to
form successively larger objects. N-body simulations of
this process originally produced fairly smooth cold dark matter (CDM) halos
\citep{1985Natur.317..595F,1985ApJ...292..371D} homogenized by poor
resolution. More recently, studies exploiting high resolution 
simulations have managed to resolve a wealth of sub-galaxy sized 
substructures (subhaloes) that can survive the violent processes associated
with merging
\citep{1998MNRAS.300..146G,1999ApJ...522...82K,1999ApJ...524L..19M,2000ApJ...539..517B,2001MNRAS.328..726S,2002MNRAS.335L..84S,2004MNRAS.355..819G,2004MNRAS.351..410G}. 

High resolution numerical simulations that resolve substructures can be
computationally expensive, and many authors choose to simulate just the
gravitational forces, modeling the baryons semi-analytically
\citep[e.g.][]{2000MNRAS.319..168C,2006MNRAS.370..645B}. Alternatively, baryons may be modeled hydrodynamically
\citep{1991ApJ...377..365K,1991ApJ...380..320N,1992ApJ...391..502K,1994MNRAS.267..401N,1997ApJS..111...73K},
with the advantage that the thermal state of the intergalactic medium
may be directly calculated. The inclusion of baryons affects the background
dark matter: \cite{1986ApJ...301...27B} showed that dissipative baryons will lead directly to the
adiabatic contraction of the halo \citep[see also][]{2004ApJ...616...16G} and
are thus a critical ingredient in determining halo and subhalo properties.

More recently a number of authors have examined these effects - gas
dissipation and star formation - on galaxy and cluster sized halos
\citep{2005ApJ...627L..17B,2009arXiv0906.0573S,2009arXiv0901.1317R}. A number 
of studies have also looked at the effect on
substructure. \cite{2005ApJ...618..557N} noted that the radial distribution of
substructures in \texttt{Adaptive Mesh Refinement} simulations was
less concentrated than both the radial distribution of subhaloes in dark matter
only runs \citep[see also][]{2004MNRAS.352..535D,2004MNRAS.355..819G} as well
as the dark matter itself. \cite{2005ApJ...618..557N} ruled out numerical
resolution as the cause showing that this bias was caused by the tidal
stripping (of dark matter) from a subhalo's 
outer regions: the bias disappears if subhaloes are chosen according to their
mass at accretion. Furthermore, \cite{2006MNRAS.366.1529M} simulated a Galactic
mass halo twice - once with baryons using \texttt{Smooth Particle Hydrodynamics} (SPH) and once with
just dark matter in order to study the effect of baryons on lensing. They
found that owing to adiabatic contraction in the 
dense cores of subhaloes, their SPH run produced an over abundance of
subhaloes in the inner regions of the halo as well as an increase by a factor
of 2 in the absolute number of subhaloes with respect to the dark matter
run. Finally, \cite{2008ApJ...678....6W} examined the effect of baryons on
substructures of group and cluster sized 
($10^{12}-10^{14}M_{\odot}$) halos with a more statistical approach. By
running the same simulation with and without gas dynamics, they found an
overall good agreement between the positions and mass function of subhaloes
across a large dynamical range with the two different techniques. Like
previous studies, they also found that the baryons steepen the 
potential well of both the host and subhaloes, making the core of the
substructure more resistant to tidal 
stripping, resulting in the relative depletion of dark matter
subhaloes towards the centre.

%YH
Baryonic substructure may effect the central DM distribution in a different
way. Satellites crossing the central part of their host halo loose energy due
to dynamical friction. This energy is being deposited in the DM component,
leading to 'heating' of the DM and consequently this may lead to the softening
of the central DM cusp \citep{2001ApJ...560..636E,2004ApJ...607L..75E}. This
has been supported by the full galaxy formation simulations of
\cite{2008ApJ...685L.105R,2009arXiv0901.1317R} and \cite{2009ApJ...697L..38J}.
%YH

\begin{figure*}
\begin{center}
\includegraphics[width=13.5pc]{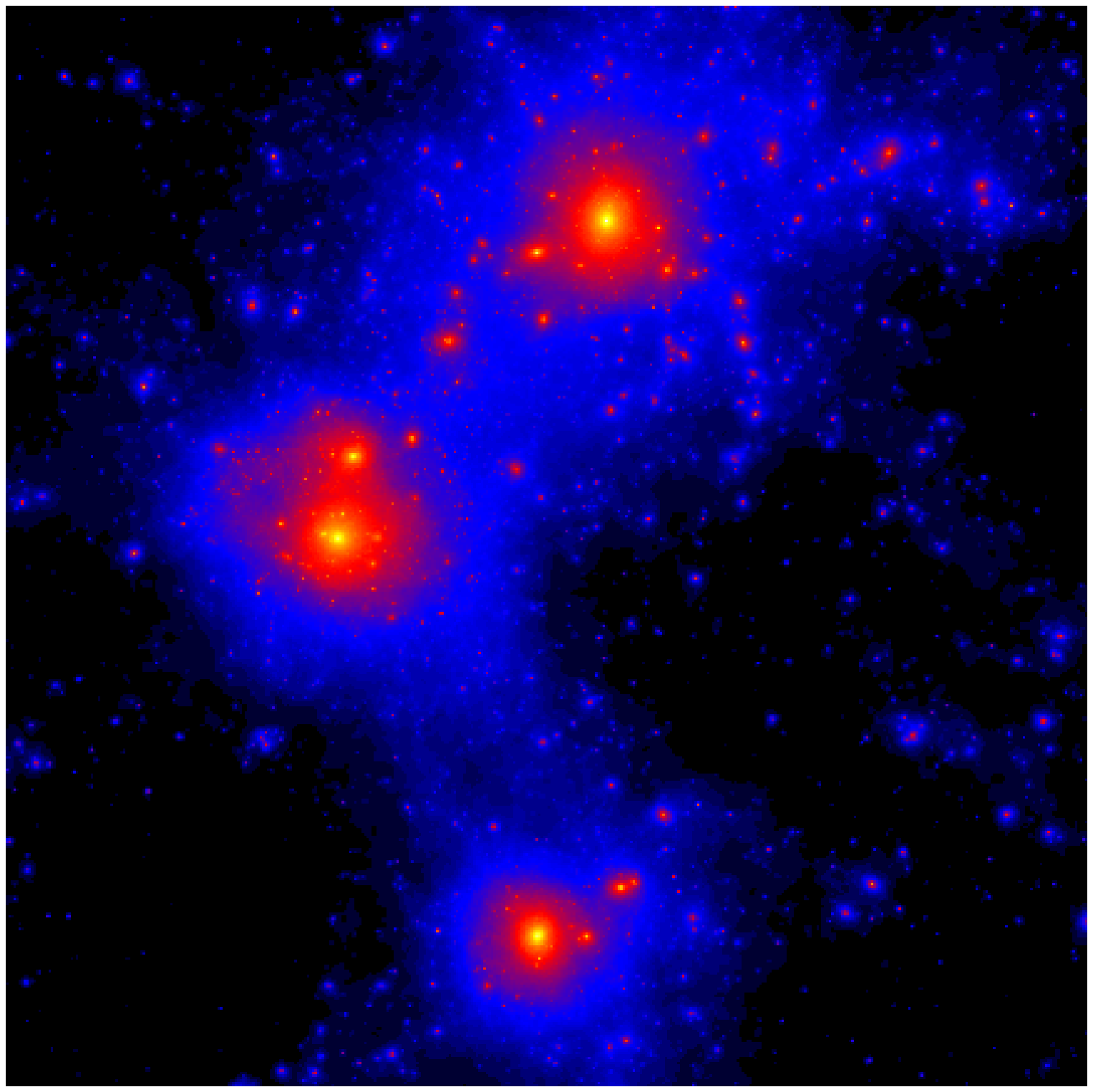}
\includegraphics[width=13.5pc]{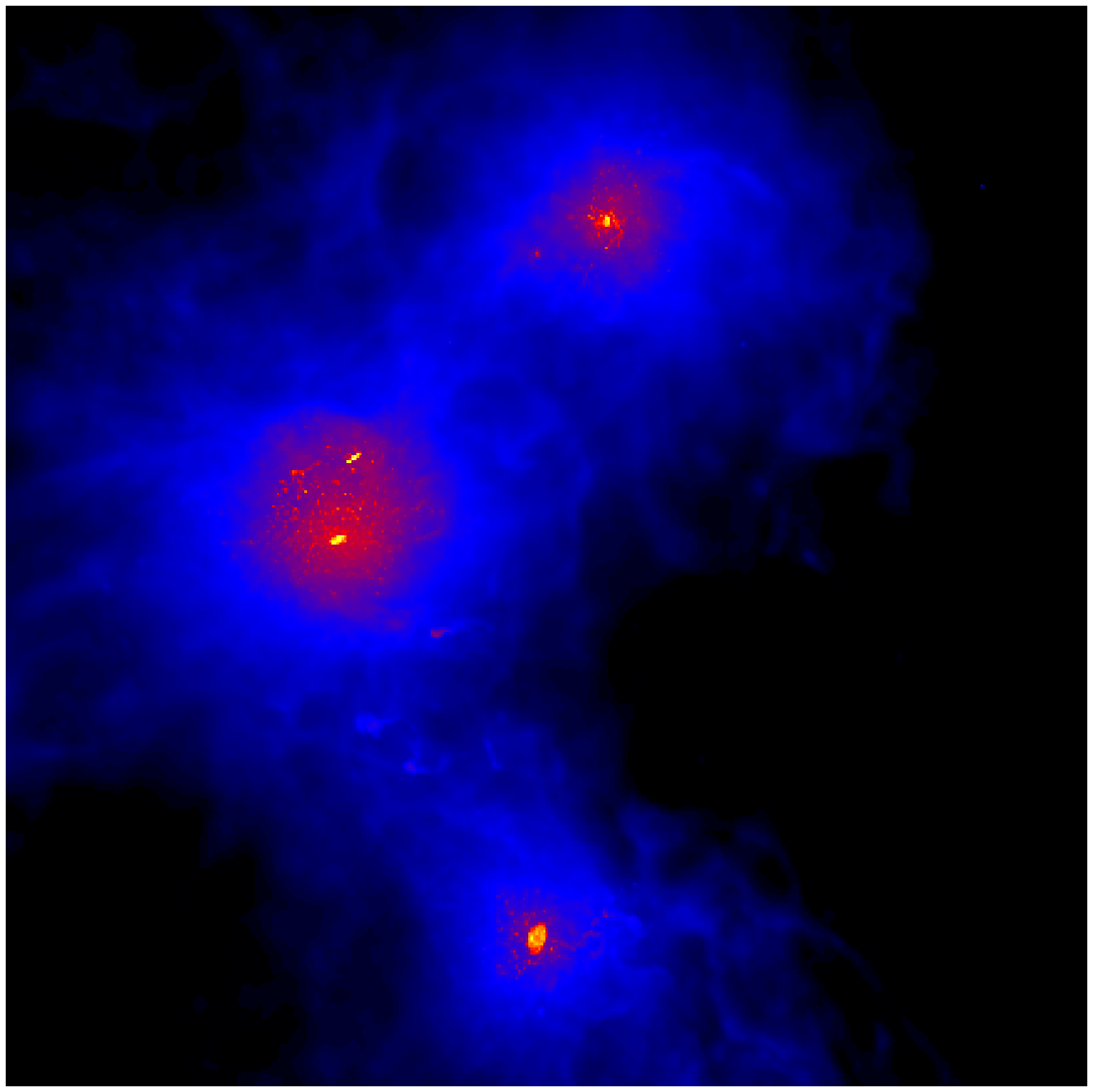}
\includegraphics[width=13.5pc]{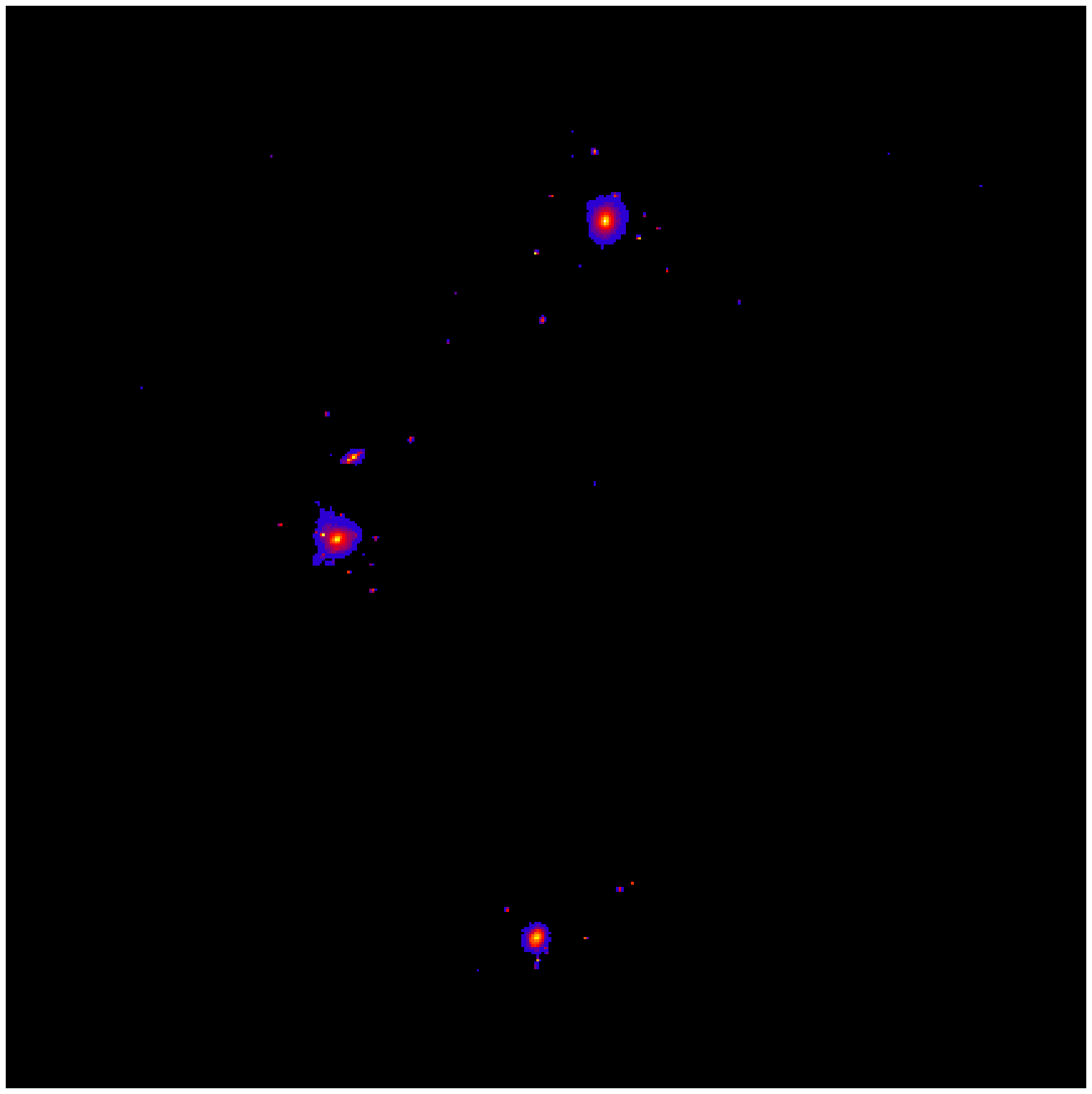}
\end{center}
\caption{The $z=0$, DM (left), gas (centre) and stellar (right) distributions
  in our simulation of the Local group. We smooth the density field at each
  point and plot the Local Group in a box with side length is 1.3 $h^{-1}
  {\rm Mpc}$. We identify as our
  ``Milky Way'' the galaxy at the top of the box, ``M31'' as the galaxy
  towards the centre and ``M33'' as the galaxy at bottom of the box. This box
  sits at roughly the correct distance from both Virgo and Coma mass clusters, as
  well as an object similar to the Great Attractor.}
\label{simpic}
\end{figure*}

One of the main new feature of the present study is that it is based on constrained
simulations of the local universe,  
where 'local' means a few tens of Mpc around the LG. 
We designed our simulation's initial conditions by imposing
constraints derived from observational data of the LG's environment;
specifically we use surveys of galaxy peculiar velocities and the
distribution of nearby X-ray selected clusters of galaxies. The
result is a simulations of a `LG' that sits within the proper dynamical
environment. This includes the correct motion and position of the Local
Supercluster, the Virgo and Coma clusters 
and the Great Attractor. The main virtue in using constrained simulations is
that it provides a numerical laboratory for testing the formation of the local
universe, and in particular the Local Group, in a numerical environment that
closely matches our actual neighborhood. Thus, much of the cosmic
scatter involved in comparing unconstrained simulations with specific
cosmological objects, is removed.

Other studies typically examine the effect of baryons in either cluster sized halos
\citep{2005ApJ...618..557N,2008ApJ...678....6W} or in a random Milky Way sized halo
\citep{2006MNRAS.366.1529M}. We have extended previous work
\citep{2005ApJ...618..557N,2006MNRAS.366.1529M,2008ApJ...678....6W} by
examining with improved mass and force resolution the the effect of baryons on
substructures in three galactic halos that resemble the Local Group.  
 
We study how the effect of dynamical friction is affected by the relative
difference in the tidal mass loss of a substructure with and without
gas-dynamics. By using two simulations - one with and one without gas dynamics -
we are able to determine where and how the
trajectories and properties of the same subhalo in the two different runs diverges. Our
simulations also have unprecedented mass and force resolution, an improvement
on previous work of between 1 and 2 orders of magnitude. We have tuned
our simulations to make it very easy to 
identify matching subhaloes in both simulations. We thus follow in
detail how the stripping of an individual subhalo is effected by the presence
of baryons and how the stripped subhalo is then affected by dynamical
friction.

This paper is organized as follows. In Sec.~\ref{meths} we describe our
methods, specifically our simulations (Sec.~\ref{sims}), our halo and subhalo
finding technique (Sec.~\ref{AHFsec}) and how we match dark matter subhaloes to
SPH ones (Sec.~\ref{sisterfindsec}). We present our results in
Sec.~\ref{results} and conclude in Sec.~\ref{concl}.

\section{Methods}
\label{meths}
In this section we describe how the simulation were constrained and run, the
method we have employed to identify substructures in each run and how we match
SPH subhaloes to DM ones.

\subsection{Initial Conditions}

In this section we describe how we generated our constrained initial
conditions. The simulations used in this paper are part of the  {\sc CLUES} (Constrained 
Local UniversE Simulations) project\footnote{{\tt http://clues-project.org}}
whose main goal is to run simulations that
mimic as accurately as permitted by current observations, the
formation of the Local Universe from cosmological initial
conditions. 

The following procedure is employed in generating constrained simulations of
the LG. We use as observational constraints peculiar velocities drawn from the
MARK III \citep{1997ApJS..109..333W}, SBF \citep{2001ApJ...546..681T} and the
local volume \citep{2004AJ....127.2031K} galaxy catalogs and the position
and virial properties of nearby X-ray selected clusters 
of galaxies \citep{2002ApJ...567..716R}. The Hoffman-Ribak algorithm \citep{1991ApJ...380L...5H} is used
to generate the initial conditions as constrained realizations of Gaussian
random fields on a $256^3$ mesh. 
These are used to perform a few low resolution DM-only N-body simulations. It
should be noted that the environmental restrictions do not constrain the
properties (e.g. position relative to eachother, mass, etc) of LG members
explicitly. Hence the low resolution simulations are searched and the one with the most
suitable LG-like object is chosen for follow up, high resolution
re-simulation. High resolution extension of the low resolution constrained
realizations are obtained by 
creating an unconstrained realization at the desired resolution,
FFT-transforming it to k-space and substituting the unconstrained low k modes
with the constrained ones. The resulting realization is made of unconstrained
high k modes and constrained low k ones. 

\subsection{The Simulations}

\label{sims}
In this section we explain details of the simulation run and, for the gas
dynamical simulation, the rules we employed for star formation and feedback. 
We choose to run our simulations using standard $\Lambda$CDM initial
conditions, that assume a WMAP3 cosmology \citep{2007ApJS..170..377S},
i.e.  $\Omega_m = 0.24$, $\Omega_{b} = 0.042$, $\Omega_{\Lambda} = 0.76$. We
use a normalization of $\sigma_8 = 0.73$ and a $n=0.95$ slope of the
power spectrum. We used the PMTree-SPH MPI code
 \texttt{GADGET2} \citep{2005MNRAS.364.1105S} to simulate the evolution of a
 cosmological box with side length of $L_{\rm box}=64 h^{-1} \rm Mpc$.

We resimulate the region of interest in our simulation according to
the following recipe. At $z=100$, we represent the linear power spectrum
by $N_{max}=4096^3$ particles of mass $m_{\rm DM} = 2.5 \times 10^{5}
h^{-1} M_{\odot} $. We use a Fourier transform of the constrained density
field to substitute the overlapping Fourier modes in to the otherwise
random realization. Following \cite{2001ApJ...554..903K}, we reduce the total number of
particles. Thus, within a lower resolution simulation of $N=256^3$
particles, we identify the position of the model local group, which
closely resembles that of the real Local Group. We then resimulate the
evolution of this region of interest, using the full resolution
(equivalent to $4096^3$ effective particles) within a sphere of just $2 h^{-1} \rm Mpc$. Outside this region, the
simulation box is populated with lower resolution (i.e higher mass) particles. We
are thus able to achieve high resolution in the region of interest, while
maintaining the correct external environment. The initial conditions were set
up by applying the Zeldovich approximation at $z=100$ to avoid spurious effects
due to cell crossing in the high resolution area.    

We use the same set of initial conditions to run two simulations, one
with dark matter only and another one with dark matter, gas dynamics, cooling,
star formation and supernovae feedback. We shall call these two simualtions DM
and SPH, respectively. In our SPH simulation, each high resolution dark
matter particle was replaced by an equal mass, gas - dark matter particle pair
with a mass ratio of roughly 1:5. In so doing we chose to identify the dark
matter particles in both runs with the same ID in order to simplify the
analysis. Numerical parameters of our simulations are summarized in Table~\ref{table1}.

\begin{table}
\begin{center}
 \begin{tabular}{l l l l l l}
      &$m_{\rm gas,i}$  &  $m_{\rm dm}$  & $\epsilon$ & $N_{\rm dm}$ \\
   \hline
   \hline
   DM: & $ - $ & $2.54  \times 10^{5}$ & $0.15$ & $5.29\times10^{7}$  \\
   
   SPH : & $4.42 \times 10^{4}$ & $2.1 \times 10^{5}$ & $0.15$ &
   $5.29\times10^{7}$  \\

    \hline

 \end{tabular}
 \end{center}
\caption{ Simulation parameters. From left to right, the initial mass per gas
  particle (in units of $h^{-1} \rm M_{\odot}$); mass of high resolution dark
  matter particles (in units of $h^{-1} \rm M_{\odot}$); softening length (in units of $h^{-1} \rm kpc$); Number of
  high resolution DM particles.}
\label{table1}
\end{table}

For the gas dynamical SPH simulation, we follow the feedback and star
formation rules of \cite{2003MNRAS.339..289S} which we briefly describe
here. The interstellar medium (ISM) is modeled  as a two phase medium composed
of hot ambient gas and  cold gas clouds in pressure equilibrium. The
thermodynamic properties of the gas are computed in the presence of a
uniform but evolving ultra-violet cosmic background generated from QSOs
and AGNs and switched on at $z=6$ \citep{1996ApJ...461...20H}. 
Cooling rates are calculated from a mixture of a primordial plasma
composition. No metal dependent cooling is
assumed, although the gas is metal enriched due to supernovae
explosions. Molecular cooling below $10^{4} {\rm K}$ is also ignored. 
Cold gas cloud formation by thermal instability, star formation, 
the evaporation of gas clouds, and the heating of
ambient gas by supernova driven winds are assumed to all
occur simultaneously.
 
For each star formation event, energy and metals are
re-injected into the ISM instantaneously. Star formation is treated in
a stochastic way and the parameters of the model are selected to 
 reproduce the Kennicut law for spiral galaxies
\citep{1983ApJ...272...54K, 1998ApJ...498..541K}. In order to prevent
excessive star formation events eminating from the same SPH particle, we allow
each gas particle to spawn at most just two generations of stars. Thus when a
star particle is formed, the mass per star is half that of the
progenitor gas particle, whose own mass is reduced accordingly.

In order to reproduce spiral disk galaxies, we assume kinetic feedback in the
form of strong winds driven by stellar explosions. We use the
stochastic approach developed by \citet{2003MNRAS.339..289S}.  SPH
particles close to the star forming regions will participate in the
wind in a probabilistic way that is proportional to the star formation rate and
the amount of supernova energy released by massive stars. We assume that a
fraction of $\beta=0.12$ of stars will explode as
supernovas.  A fraction of this energy is used to energize (kick)
particles in the high density regions surrounging sites of star formation. This
removes low angular momentum gas from the center of dark halos,
and, by inhibiting the overcooling problem, contributes to the formation of
extended gaseous disk. 

We have tested this mechanism and found it to be vital in the production of
stable disks containing both gas and 
stars in our simulated halos. The same simuations run with only thermal
feedback and without winds, are not able to create disks at all - rather the
baryons and end up as a feautureless spheroid of gas and stars
\citep[e.g. see][and references therein]{2005MNRAS.363.1299O}. We note, that
the size (extent) of the simulated disks are related to the resolution of our
simulations. We have run a second simulation from the same initial
conditions and with identical star formation and feedback rules but with lower
mass resolution ($2048^3$ instead of $4096^3$ effective 
particles: that is, with particles that are 8 times more massive). In this low
resolution run, we find that the stellar disks that were produced were thicker
and smaller. A detailed analysis of the results from these simulations will be 
reported in Yepes et al 2009 (in prep).

\begin{table}
\begin{center}
 \begin{tabular}{l l l l c l}
% \begin{tabular}{l p{9mm} l p{9mm} l p{9mm} l p{9mm} l p{9mm} l p{9mm}}
      &M$_{\rm tot}$ & r$_{\rm vir}$ & N$_{p}$  & $\frac{\rm M_{\rm
   b}}{\rm M_{\rm tot}}$ & N$_{\rm sats}$  \\
   \hline
   \hline
   DM: MW : & 6.57 & 230.84 & 1808918  & - & 13  \\
       M31 : & 8.17 & 248.26 & 2249954 & - & 29 \\
       M33 : & 3.21 & 181.96 & 885833 & - & 7\\
    \hline
   SPH: MW : & 5.71 & 220.40 & 1745996  & 0.08 & 20  \\
        M31 : & 7.81 & 244.58 & 2288039 & 0.12 & 26 \\
        M33 : & 2.90 & 175.71 & 856820 & 0.11 & 8\\
   \hline

 \end{tabular}
 \end{center}
\caption{The properties of the three local group objects, in the DM (top) and
   SPH (bottom) simulations. The colums show, from left to right, (1) M$_{\rm
   tot}$, the total virial mass in units of $10^{11}M_{\odot}$, (2) r$_{\rm vir}$,
   the virial radius in units of kpc, (3) N$_{p}$, the total number of DM particles
   within the virial radius, (4) $\frac{M_{\rm b}}{M_{\rm tot}}$, the
   baryon fraction within the virial radius, and (5) N$_{\rm sat}$, the number
   of satellites at $z=0$ above our $2\times 10^{8} M_{\odot}$ mass cut.}
\label{table2}
\end{table}

The constrained simulations studied here are zoomed on a LG-like object whose
three main galaxies - the Milky
Way (MW), M31 and M33 - were reproduced with masses in good agreement with the
latest mass estimates of these objects as reported in the literature
\citep[see][and references
therein]{2002ApJ...573..597K,2008ApJ...684.1143X,2008MNRAS.389.1911S}. In
Table~\ref{table2} we give the characteristic properties of each local group halo
for both our DM and SPH run. Note that the halos' virial radii (and therfore
their virial masses) are not identical in the two runs. This is due to the
adiabatic contraction of the baryons in the SPH run which results in a denser core and
therefore a smaller virial radius. Moreover, the total mass is reduced due to
the lower baryon fraction compared with the cosmic mean. If we choose to calculate the mass within
some fixed radius in both simulations (e.g. the average virial radius) and
correct for the reduced baryon fraction, we obtain halo masses in the two
simulations consistent with eachother to within a few percent. 

\begin{figure}
\begin{center}
\includegraphics[width=20pc]{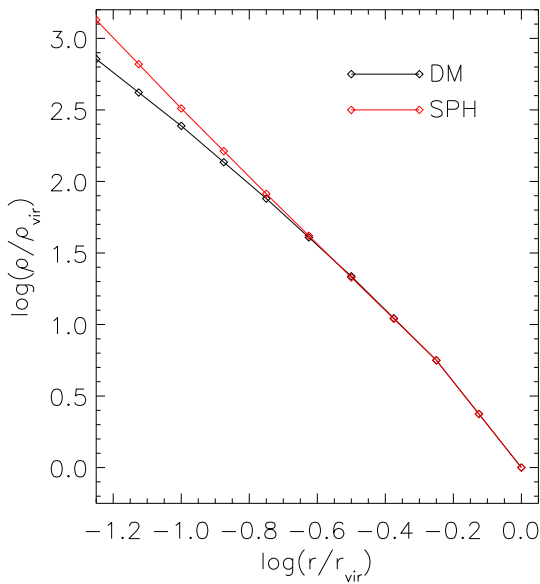}
\end{center}
\caption{The average total matter density profile of our three halos in our DM (black) and SPH (red) simulations, normalized to the virial radius and virial over-density. Note that these are the same across much of the halo, differing only for the innermost regions.}
\label{densprof}
\end{figure}

In Fig.~\ref{simpic}, we plot the matter
distribution of the three main components of our local group SPH
simulation. We plot the smoothed density of the dark matter, the gas and the
stars. It is interesting to note that the dark matter distribution in the DM run is 
the same as the dark matter distribution in the SPH run for the entire halo except for the inner most regions. Only within around $ 0.01~r_{\rm vir}$ does the condensation of baryons in the halo's center  significantly alter the underlying distribution of dark matter particles.  The total matter distribution in the two runs however, differs on a slightly larger scale. In Fig.~\ref{densprof} we show the mean interior density profile averaged over our three halos in our two simulations. Across most of the halo's radius the density profiles are the same: only within $\sim 0.1~r_{\rm vir}$ does the total density profile in the SPH simulation rise significantly above the DM one\footnote{We note however, that in the analysis that follows few of our subhaloes penetrate the region, $< \sim 0.1~r_{\rm vir}$, where the potential is significantly different in the SPH run.}.

\subsection{Identification of Halos and Subhaloes}
\label{AHFsec}
In this section, we explain how our halo and subhalo finding algorithm works.
In order to identify halos and subhaloes in our simulation we have run the
MPI+OpenMP hybrid halo finder \texttt{AHF} (\texttt{AMIGA} halo finder, to be
downloaded freely from \texttt{http://popia.ft.uam.es/AMIGA}) described in
detail in \cite{2009ApJS..182..608K}. \texttt{AHF} is an improvement of the
\texttt{MHF} halo finder \citep{2004MNRAS.351..399G}, which locates local
over-densities in an adaptively smoothed density field as prospective halo
centers. The local potential minima are computed for each of these density
peaks and the gravitationally bound particles are determined. Only peaks with
at least 20 bound particles are considered as haloes and retained for further
analysis. We would like to stress that our halo finding algorithm automatically
identifies haloes, sub-haloes, sub-subhaloes, etc. For more details on the
mode of operation and actual functionality we refer the reader to the
code description paper \citep{2009ApJS..182..608K}. 

For each halo, we compute the virial radius $r_{\rm vir}$, that is the radius
$r$ at which the density $M(<r)/(4\pi r^3/3)$ drops below $\Delta_{\rm
  vir}\rho_{\rm back}$. Here $\rho_{\rm back}$ is the cosmological background
matter density. The threshold $\Delta_{\rm vir}$ is computed using the spherical
top-hat collapse model and is a function of both cosmological model and
time. For the cosmology that we are using, $\Delta_{\rm vir}=355$ at $z=0$.  

Subhaloes are defined as haloes which lie within the virial radius of a more
massive halo, the so-called host halo. As subhaloes are embedded within the
density of their respective host halo, their own density profile usually
shows a characteristic upturn at a radius $r_t \lsim r_{\rm vir}$, where
$r_{\rm vir}$ would be their actual (virial) radius if they were found in
isolation.\footnote{Please note that the actual density profile of subhaloes
  after the removal of the host's background drops faster than for isolated
  haloes \citep[e.g.][]{2004ApJ...608..663K}; only when measured within the
  background still present will we find the characteristic upturn used here to
  define the truncation radius $r_t$.}  We use this ``truncation radius''
$r_t$ as the outer edge of the subhalo and hence subhalo properties
(i.e. mass, density profile, velocity dispersion, rotation curve) are
calculated using the gravitationally bound particles inside the truncation
radius $r_t$. For a host halo we calculate properties using the virial radius $r_{\rm vir}$.

We build merger trees by cross-correlating haloes in consecutive simulation
outputs. For this purpose, we use a tool that comes with the \texttt{AHF}
package and is called \texttt{MergerTree}. As the name suggests, it serves the
purpose of identifying corresponding objects in the same simulation at different
redshifts. We follow each halo (either host or subhalo) identified at redshift
$z=0$ backwards in time, identifying as the main progenitor (at the previous
redshift) the halo that both shares the most particles with the present halo \textit{and} is
closest in mass (see Eq.~\ref{merit}). The latter criterion is important for subhaloes given that all their
particles are also typically bound to the host halo, which is typically orders
of magnitude more massive. Given the capabilities of our halo
finder \texttt{AHF} and the appropriate construction of a merger tree,
subhaloes will be followed correctly along their orbits within the
environment of their respective host until the point where they either are
tidally destroyed or directly merge with the host.

\subsection{Matching Halos in the two simulations}
\label{sisterfindsec}
In this section we describe how we match subhaloes in our SPH simulation to DM
``sister'' subhaloes. Our simulations are set up such that the identifiers (ids) of dark matter
particles are the same in both simulation. This allows us to locate
which SPH subhalo corresponds to a specific DM subhalo. For a
given SPH subhalo we locate the DM subhalo which contains both the greatest
fraction of its dark matter particles and which is closest to it in mass.
We call this DM halo, the SPH subhalo's ``sister''. Specifically, for
each dark matter halo candidate $i$ we calculate a merit $M_{i}$, defined as 
\begin{equation}
  M_{i}=\frac{n^{2}_{\rm shared}}{n_{\rm dm, \it i}n_{\rm sph, \it i}}
\label{merit}
\end{equation}
where $n_{\rm shared}$ is the number of particles found in both subhaloes, $n_{\rm
  dm,\it i}$ is the number of particles in the DM sister candidate and $n_{\rm sph,\it i}$ is
  the number of particles in the SPH subhalo. We then identify the sister
  as that halo with the largest value of $M_{i}$. Typically, successfully
  matches result in $M_{i} \gsim 0.5$ and we assume to have failed to find a dark
  matter sister if $M_{i} < 0.2$ for a given sph subhalo.

\begin{figure}
\begin{center}
\includegraphics[scale=0.195]{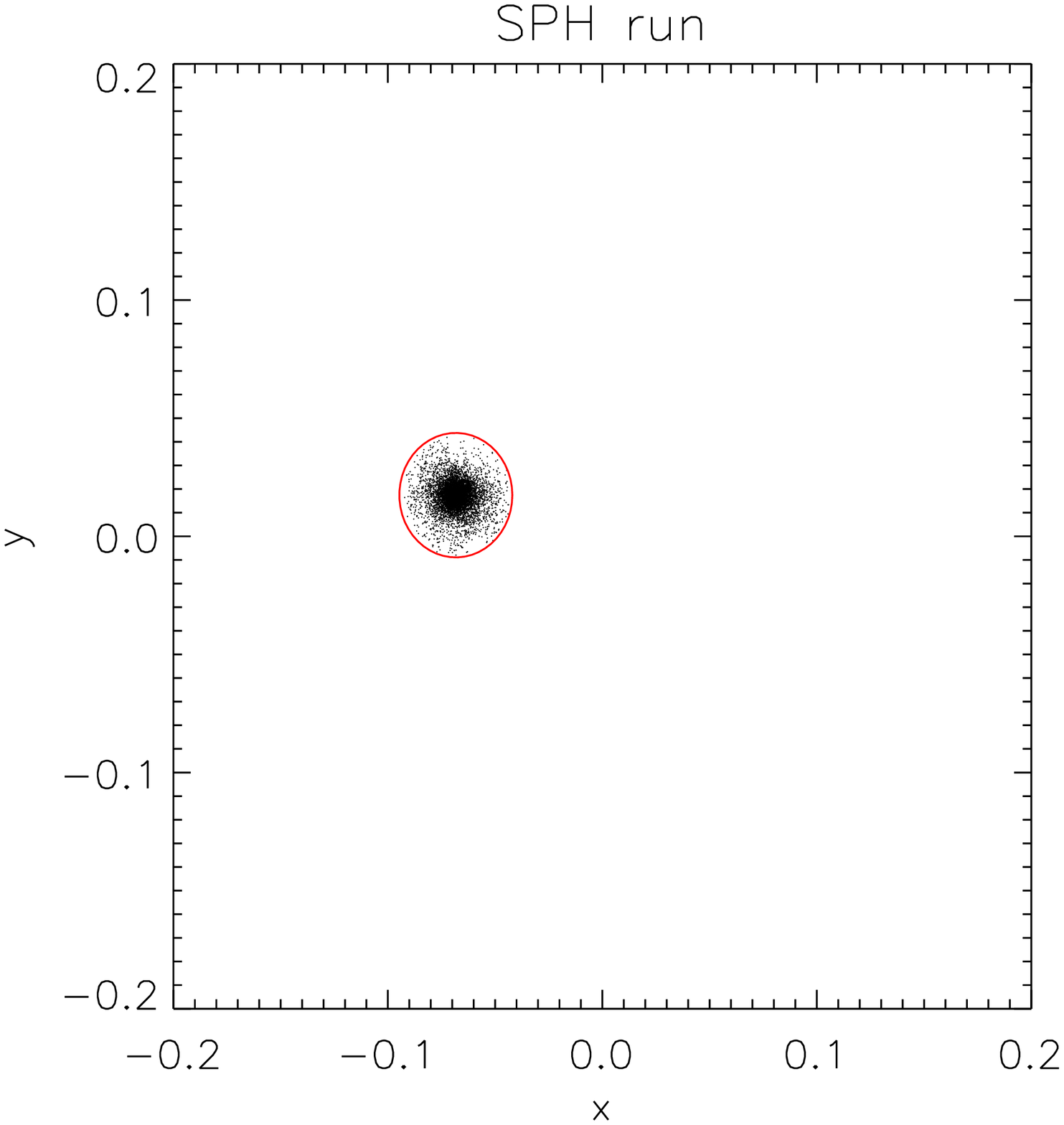}
\includegraphics[scale=0.195]{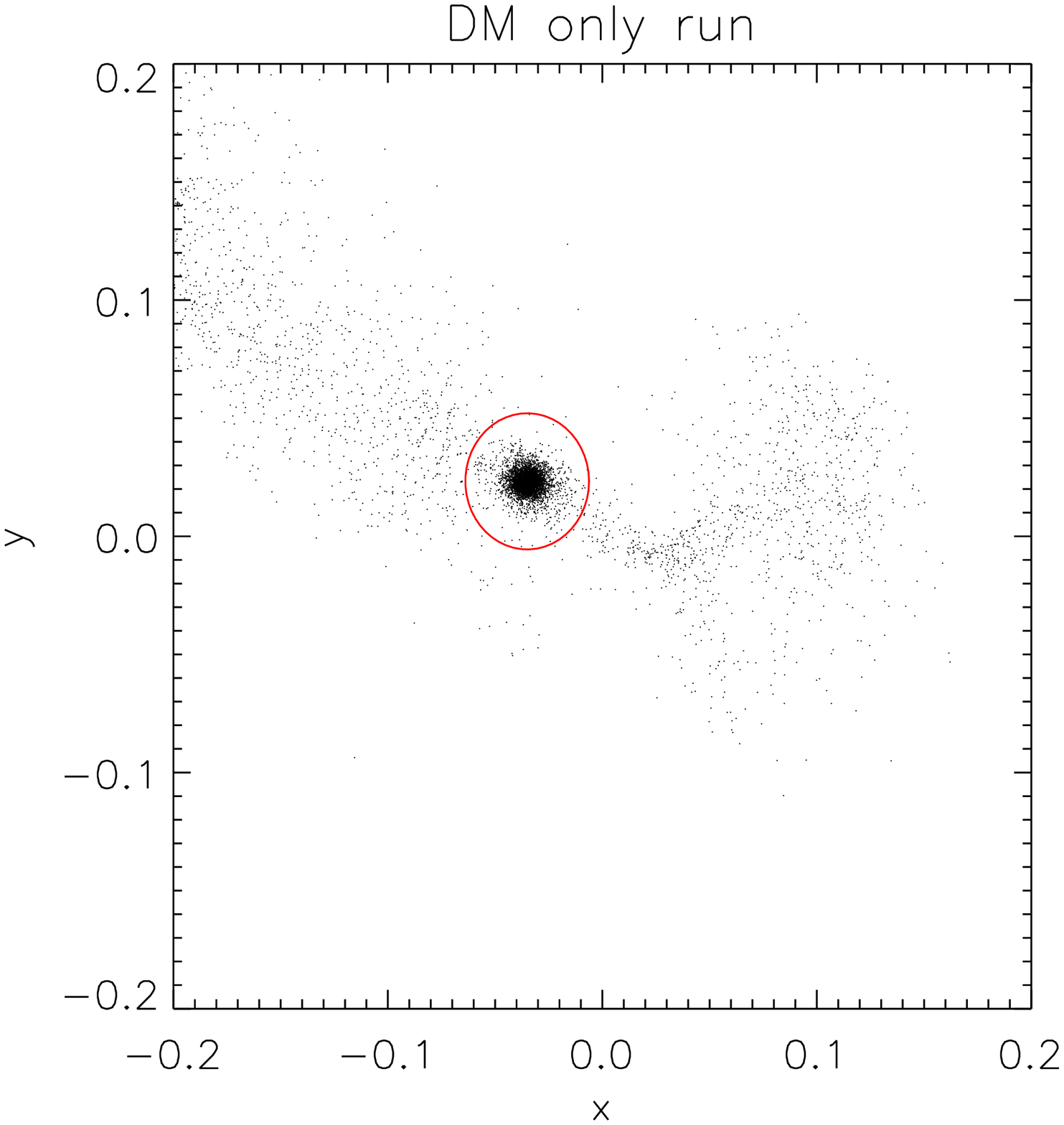}
\end{center}
\caption{How subhaloes in the two different simulations are matched. On the
  left side we show all the dark matter particles belonging to a specific
  subhalo in the SPH
  simulation, in the frame of the host halo. The background dark matter, within which the subhalo is
  embedded,  has been omitted for clarity. The red circle indicates the subhaloes radial extent. On the
  right, we show the same set of particles in the DM run. Note that 
  they define an extended object. In red we indicate the DM subhalo to
  which the SPH subhalo was matched. Again, the background dark matter is
  omitted. Note that the DM (sister) subhalo contains additional dark matter particles -
  not found in the SPH subhalo and thus not plotted here - that are none
  the less bound to the DM substructure.} 
\label{match}
\end{figure}
We demonstrate how our sister finding algorithm works in Fig.~\ref{match} for
one specific subhalo. The dark matter particles in the DM run with
the same ids as the dark matter particles belonging to an SPH
subhalo are found extended in the DM run. However, despite being
extended, often enough of these particles clump together in one single
subhalo, albeit with other dark matter particles \textit{not} found in the SPH
subhalo. We take as the DM ``sister'', the \textit{entire} subhalo that shares
the most particles with the SPH subhalo.

If we wanted to, we could have tried to find sisters by doing the reverse:
locating particles that belong to a substructure in the DM simulation, in the
SPH simulation, and finding which SPH subhalo hosts the most of them. The
resulting matches are equivalent.

It is interesting to note that our failure to match an SPH subhalo to a DM one
is not a strong function of radius. Although more such failures to match occur
at smaller radii of $\sim r<0.1~r_{\rm vir}$ \citep[owing to the complete destruction of the DM subhalo,
i.e.][]{2005ApJ...618..557N}, beyond $\sim 0.1~r_{\rm vir}$ the
distribution of Merit values (i.e. Eq.~\ref{merit}) is uniform. This implies
that although the complete
destruction of a DM subhalo may be responsible for the steepening of the
density profile of substructure in the inner most parts (i.e. within $r < 0.1r_{\rm vir}$), it cannot explain the
fact that SPH subhaloes are more radially concentrated across the halo's full
virial radius.

\section{Results}
\label{results}
%\subsection{Differences in the global properties of halos}
%\input{global_sec.tex}
%\subsection{The effect of baryons on the $z=0$ distribution of substructure}
In this section we explain how baryons effect the properties of substructures
in our two simulations. Since, as mentioned in Sec.~\ref{AHFsec}, we catalogue
all bound objects composed of more than 20 particles, the vast majority of
substructures in each of the three 
local group objects, have very low mass. Since particle mass is not constant in
SPH runs (due to star formation and feedback processes) it is incorrect to
compare the full sub-halo population found in our DM simulation with that
found in the SPH run. In order to compare like with like, we make a mass cut, comparing only
the most massive of the $z=0$ subhaloes. We select subhaloes whose mass is
$M_{\rm sub} > 2\times 10^{8}h^{-1} M_{\odot}$, which roughly corresponds to
subhaloes with more than 1000 particles. A total of around $\sim 50$ subhaloes
of our three main objects meet this cut in both simulations. This mass cut was
chosen to obtain roughly the same number of satellites as seen orbitting about
the Milky Way.

\begin{figure}
\begin{center}
\includegraphics[width=20pc]{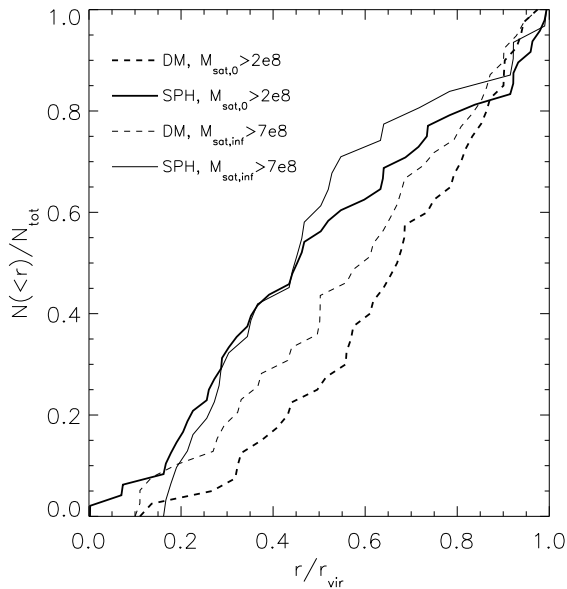}
\end{center}
\caption{The cumulative radial density profile of substructures at $z=0$. The
  thick solid and dashed line, shows this quantity for SPH and DM subhaloes
  (respectively) when considering a $z=0$ mass 
  cut of $M_{\rm sat, 0} > 2\times 10^{8}h^{-1} M_{\odot}$. The thin solid and
  dashed line show this quantity for SPH and DM subhaloes when considering only
  subhaloes that were more massive then $M_{\rm sat, inf} > 7\times 10^{8}h^{-1}
  M_{\odot}$ at infall. The infall mass cut was
  chosen in order to have the same number of objects as the $z=0$ cut.}
\label{cumdenprof}
\end{figure}

In Fig.~\ref{cumdenprof} we show the cumulative radial profile of substructures
in our two simulations at $z=0$. We make two different mass cuts, one
considering the $z=0$ mass of a substructure, and a second
\citep[following][]{2005ApJ...618..557N} which considers the subhalo's mass at
infall. In order to compare the same number of satellites, we consider the
$z=0$ position of only those
satellites which at infall where more massive then $M_{\rm sub} > 7\times
10^{8}h^{-1} M_{\odot}$.

What is strikingly apparent from Fig.~\ref{cumdenprof}, 
is that the SPH substructures are significantly more radially
concentrated than the DM ones. This feature is irrespective of which mass cut
we look at, although the trend is stronger when considering the $z=0$ mass
cut. For example, within 0.5 $r_{\rm vir}$
we find $\sim 55 \%$ of SPH substructures compared with just $\sim
35\%$ DM ones. This result in consistent with that of
\cite{2008ApJ...678....6W}, who compared the radial distribution of much
more massive substructures ($M_{\rm sub} \sim 10^{10} M_{\odot}$) in more
massive systems ($M_{\rm host} \sim 10^{12-14} M_{\odot}$) in dark matter only
and SPH simulations of galaxy groups. 

The tendency for SPH subhaloes to be more radially concentrated at
$z=0$ weakens with increasing redshift. Beyond $z=0.75$, the presence of baryons
has little discernible effect on the radial concentration of subhaloes and the
DM and SPH simulations return subhaloes with nearly identical radial
distributions. Indeed, as we go to higher redshifts we find that all SPH and
DM subhaloes are less radially concentrated. 

\begin{figure}
\begin{center}
\includegraphics[width=20pc]{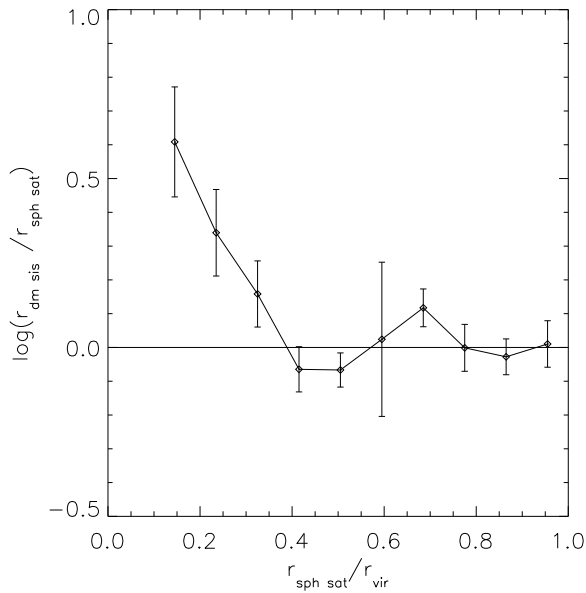}
\end{center}
\caption{The $z=0$ difference in the radial position of SPH satellites and
  their DM sisters plotted versus the radial position of the SPH
  satellite. }
\label{radif}
\end{figure}

Fig.~\ref{cumdenprof} indicates that SPH subhaloes are more
centrally concentrated than DM ones. Since we are able to
match subhaloes in both simulations, we can examine the difference in
radial position of an individual DM - SPH subhalo pair. In
Fig.~\ref{radif} we show the difference in the radial position of two sister
halos as a function of the SPH subhalo's radial position.
By examining how the difference in position changes with radius,
Fig.~\ref{radif} indicates that the higher radial concentration of SPH
subhaloes seen in Fig.~\ref{cumdenprof} is due to a relative depletion of
DM sisters in the inner parts of the halo. Beyond 0.4 $r_{\rm vir}$
DM sisters and SPH subhaloes are located at roughly the same radial
distances. Within this radius however, DM sisters have not dissapeared (by
falling below the mass cut). Rather, they are simply located up to 3 times
further out, for SPH subhaloes located at $\sim 0.15 r_{\rm vir}$, and up to 1.5
times further out for those located at $\sim 0.3 r_{\rm vir}$. 

The scarcity of DM subhaloes in the inner parts of the host
halo is due to the subtle changes to the substructures potential well caused
by the presence of baryons. In particular, the deepening of the potential well
due to a cold baryonic component will directly impact the two main
physical processes affecting a substructure's dynamics: dynamical friction and
tidal stripping of substructure material. 

The dynamical friction time scale goes as $\tau_{\rm dyn}\sim
\frac{1}{M_{\rm sat}}$, \citep[e.g.][]{1975ApJ...202L.113O,2009PhT....62e..56B} implying that the most massive
subhaloes are those whose orbit's should lose the most energy and are therefore
most likely to in spiral and be found in the inner parts of the host. Yet
tidal stripping causes $M_{\rm sat}$ to decrease, making the satellite less
susceptible to the dynamical friction.

\begin{figure}
\begin{center}
\includegraphics[width=20pc]{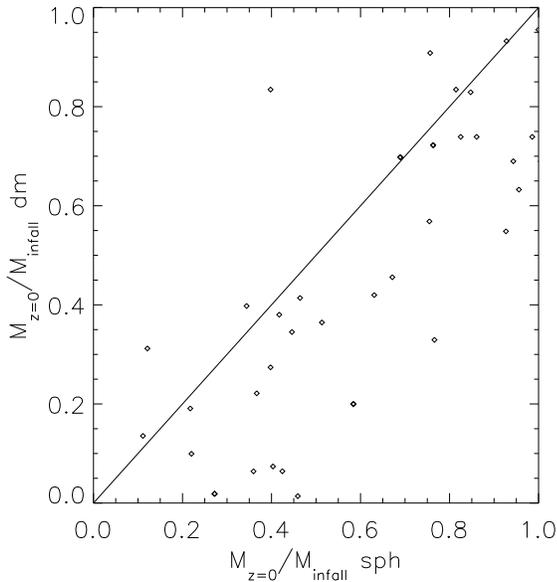}
\end{center}
\caption{The fraction of infall mass retained at $z=0$ for subhaloes identified
in SPH versus their matched DM sisters. Objects
that sit on the diagonal line have lost the exact same amount of mass in both
runs. Those objects in the lower triangle represent subhaloes which have lost
more mass in the DM run while those objects in the upper triangular
half have lost more mass in the SPH run. }
\label{mloss}
\end{figure}

The amount of material stripped can be quantified by looking at the ratio of
the mass retained at $z=0$ to the mass a satellite had when it was
accreted. In Fig.~\ref{mloss} we plot this ratio for SPH satellites versus the
matched DM sisters. Rather than sitting directly on
the diagonal line indicating equal amounts of tidal stripping, nearly all DM -
SPH subhalo pairs sit in the triangular lower half of Fig.~\ref{mloss}
indicating that the DM sister lost more mass than the SPH subhalo since accretion. In some cases the difference is extreme, with
subhaloes in SPH losing half their mass and their dark matter sister losing
99 \% of their infall mass by $z=0$. Although a handful of SPH subhaloes
have retained less than their dark matter sisters, the fraction of all
subhaloes for which this is true is small ($\sim 5\%$) and the relative mass
loss is not as extreme.

We have seen from Fig.~\ref{mloss} that DM subhaloes are more likely to
retain less of their infall mass by $z=0$. Yet if our hypothesis is correct -
that those DM subhaloes that have suffered the most tidal stripping have
consequently suffered less dynamical friction - we should expect to see a
correlation between the amount of mass retained and the position in the halo
at $z=0$. Indeed, those SPH halos that manage to fall
into the centre of the halo - leaving their DM sisters behind in the outer
regions of the halo - are also those whose DM sisters have lost the most
mass. In other words, when a subhalo falls into a host halo, if it falls in
early enough and if its orbit is such that it will lose mass due to tidal stripping, the stripping of the DM will decrease the dynamical friction
on the DM subhalo, allowing the SPH subhalo to penetrate to the centre while
leaving the DM subhalo behind. This notion can be tested by looking at the satellite's energy as a function of time. When comparing the energy loss of a SPH-DM satellite pair, we find that the DM subhalo has indeed lost more orbital energy, although the relative difference in energy loss is moderate ($\sim 10\%$).

\begin{figure}
\begin{center}
\includegraphics[width=20pc]{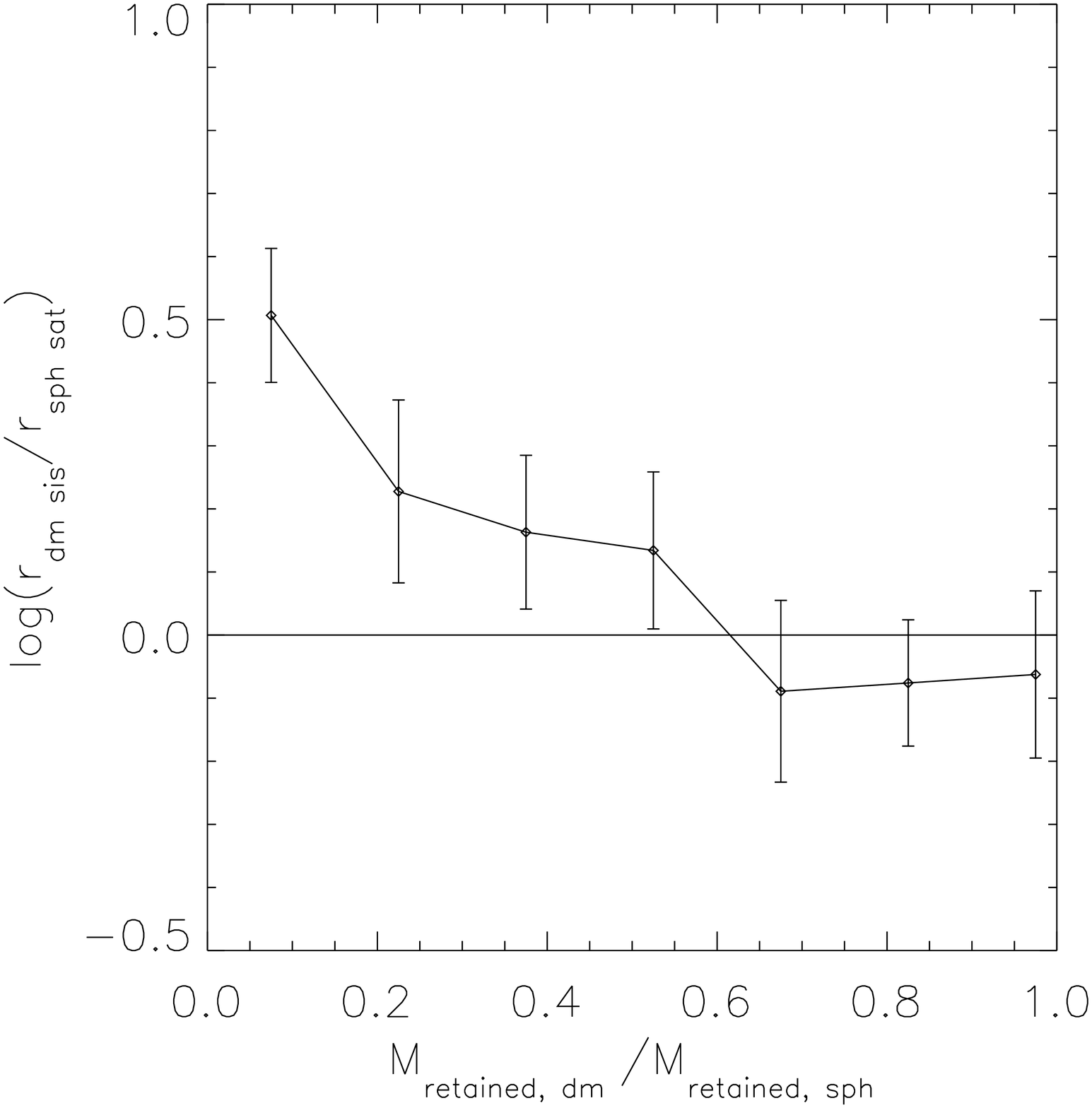}
\end{center}
\caption{ The ratio of mass retained at $z=0$ in a DM subhalo to that
  retained in an SPH halo plotted versus the $z=0$ difference in radial
  position of the pair.}
\label{mlr}
\end{figure}
We can see this behavior explicitly in Fig.~\ref{mlr} where we plot the
difference in the $z=0$ radial position between SPH and DM subhaloes
as a function of the ratio of the infall mass that is retained by each subhalo
$z=0$. It is interesting to note that the $x$-axis of Fig.~\ref{mlr} extends to
$\sim 1$ implying that very few SPH subhaloes lose more mass then their DM
sisters. The amount of mass lost by a DM subhalo is well
correlated with the difference in $z=0$ radial position, with pairs that are
close together having lost similar amounts of mass, while for those spread far
apart (i.e. where $r_{\rm dm} \gsim 1.5~r_{\rm sph}$) indicate that the DM
subhalo has retained only $\lsim 30 \%$ of the mass retained by the SPH subhalo.

\begin{figure}
\begin{center}
\includegraphics[width=20pc]{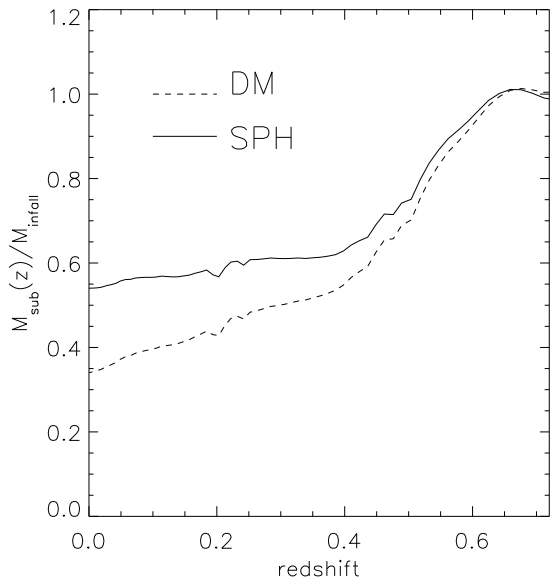}
\end{center}
\caption{The mass of an SPH subhalo (solid line) and its DM sister (dashed line) as a
  function of redshift, normalized to the mass each subhalo had at infall,
  around $z=0.72$. Note how the masses really start to deviate after infall,
  with the DM subhalo retaining only $\sim 35\%$ of its infall mass,
  while the SPH satellite retains $\sim 55\%$, a factor 1.5 more. }
\label{masshist}
\end{figure}

\begin{figure}
\begin{center}
\includegraphics[width=20pc]{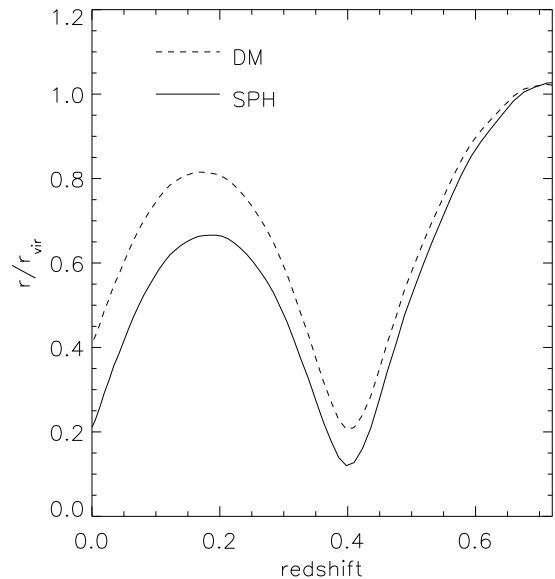}
\end{center}
\caption{The distance from the main halo of an SPH subhalo (solid line) and
  its DM sister (dashed line) as a
  function of redshift. Note how the trajectories start to diverge shortly
  after infall. The DM subhalo is consistently further away from the centre of
  the main halo then the SPH subhalo.}
\label{disthist}
\end{figure}

We now follow in detail one specific SPH - DM sister subhalo pair, in order to
highlight the salient aspects that are effected by tidal stripping and
dynamical friction. We have specifically chosen a subhalo which can be used as
an exemplary illustration for this purpose. The mass of this subhalo, $M_{\rm sub} \approx
10^{9}h^{-1}M_{\odot}$ is typical of satellites in our sample, and was accreted at $z=0.7$, an unexceptional redshift. In Fig.~\ref{masshist} we show the mass of this subhalo as
function of redshift (since accretion) onto the main halo. The DM
sister (dashed line) closely follows the SPH subhalo's mass for a short
period directly after accretion. However, by $z=0.4$ the DM halo has
already lost a good $\sim10\%$ more then its SPH counterpart. Although the
subhalo in both runs loses mass due to tidal stripping, by $z=0$ the DM sister
has lost $\sim65\%$ of its infall mass, while the SPH subhalo
has only lost $\sim 45\%$ of its infall mass.

We may compare the mass loss presented in Fig.~\ref{masshist} with the radial
distance from the centre of the main halo. In Fig.~\ref{disthist} we show the
orbit of this subhalo as a function of redshift. Although the orbits at infall
are identical, they start to diverge shortly thereafter. By the first
pericentric passage, the SPH subhalo manages to penetrate further in (by about
a factor of 2) than the DM sister. Similarly, at apocentre the DM halo is
$\sim 0.15~r_{\rm vir}$ further away then the SPH subhalo. The DM
sister is consistently further behind the SPH subhalo at all stages of the
orbit, resulting in the SPH subhalo being nearly a factor of 2 closer then the
DM subhalo at $z=0$.

Tidal stripping is most likely to dislodge loosely bound material on a
subhaloes periphery, where any change in the host's potential can have the
greatest effect. \cite{2008ApJ...673..226P} showed that tidal debris fields that pre
and succeed a subhalo on its orbit are composed mostly of particles depleted
from this peripheral region. Since the cold baryons in our subhaloes have all
collapsed to the centre of the clump's potential before accretion, we except that the
stripped material will be composed of a mixture of DM and hot gas
particles, loosely bound to the subhalo. Indeed the baryon
fraction of this subhalo increases as material is stripped from it. By $z=0$
the baryon fraction has risen by a factor of more than three compared with its
baryon fraction at infall (not explicietely shown here). Since the total mass of the subhalo has decreased yet the
baryon fraction has increased, we are led to the conclusion that the stripped
material was mostly DM.

\section{Conclusions and Discussion}
\label{concl}
In this paper, we have examined the effect of dissipative baryons -  treated
hydrodynamically with sophisticated rules for star formation and feedback - on
halo substructure. We have extended the work of previous authors
\citep{2005ApJ...618..557N,2006MNRAS.366.1529M,2008ApJ...678....6W} to include
not just a statistical description of the entire subhalo population, but a
detailed study of how a single SPH subhalo behaves differently than its DM
counterpart. 
Our simulations bring a new approach to the study of the LG and the nearby universe. 
The simulations are constrained to reproduce the local environment, thereby enabling the testing of the 
processes of galaxy formation in a controlled environment that closely
resembles the observed nearby universe. The present simulations also surpass previous studies in mass and force resolution.

We use our simulations to examine the
radial distribution of subhaloes. We find that subhaloes located in the inner
parts of thier hosts and above a certain mass cut are more concentrated at $z=0$ in the SPH simulation than in the DM one. The difference in radial distribution between DM and SPH subhaloes above a
certain mass at a given epoch, decreases with higher redshift, such that by
z=0.75 the two populations have similar distributions. The difference in radial distribution weakens if
subhaloes are selected according to infall mass (rather than $z=0$ mass) but
it does not - as suggested by \cite{2005ApJ...618..557N} who looked at clusters
3 orders of magnitude more massive - completely disappear.  

The relative suppression of dark matter subhaloes located in the central
regions of a host, is in part due to their susceptibility to tidal
forces. As suggested by \cite{2005ApJ...618..557N}, the disintegration of DM subhaloes in the inner parts of the host halo is in part responsible for their reduced numbers there. SPH subhaloes, with cold dense baryons in their centres, are able to inhibit the amount of material tidally stripped and
thus hold on to more material. Any material that is stripped off an SPH subhalo,
tends to originate in the outer regions of the subhalo and is therefore mostly
composed of dark matter - little baryonic material is stripped.

Since DM subhaloes lose more mass, they are also less likely to lose orbital
energy due to dynamical friction. Our results suggest that although the DM subhaloes lose more
mass due to tidal stripping, the (relatively) more massive SPH subhaloes spiral into the central region of the host. For a DM-SPH subhalo pair, this results in widely separated subhaloes.

However, if tidal stripping is not particularly effective, dark matter subhaloes trace their SPH sisters. This principally occurs when subhaloes are orbiting in the outer parts of the halo. In the inner parts, the SPH subhaloes are typically closer to the centre of the host than their dark matter sisters. The final $z=0$ masses of two sisters can differ wildly due to
the vastly different histories each subhalo has had since accretion. Dark matter
subhaloes almost always lose more mass than an SPH one, reflecting their more
feeble nature. 

Our results are important for any study of substructures in dark matter only
simulations that focus on the inner parts of the host halos. For
example, \cite{2009arXiv0903.4559X} performed an indepth study on the
anomalous flux ratios observed with strong lensing of background QSOs. Using
the Aquarius simulation they found that too few subhaloes exist in the central
regions to account for the observed anomalies. Yet they qualify their study
by explicitely stating that the increased survivability of subhaloes due to
the presence of baryons will have an unknown effect on their results. Our work
indicates that the effect of baryons may be important for massive subhalos
that have small pericentric distances.

Our results are important for semi-analytical models of galaxy formation which
use merger trees extracted from pure dark matter simulations to calculate the
properties of satellite galaxies \citep[e.g.][]{2005MNRAS.363..146L}. In these cases, dark matter subhaloes are often destroyed by tides earlier then they would be if hydrodynamics were also included. Thus, it is important to bear in
mind that the mass of a DM subhalo located towards the centre, will be
routinely underestimated.  Any semi-analytical model that uses the mass of a subhalo extracted from a dark matter simulation to make predictions on the properties of the satellite galaxy it hosts, will suffer from this systematic effect.

The results of \cite{2005ApJ...618..557N} indicate that the bias towards more radially concentrated subhalo distributions disappears when subhaloes are selected according to any
property that is unaffected by the processes associated with accretion, such
as their maximum circular velocity. We test this hypothesis by looking at the
radial distributions of subhaloes chosen according to infall mass (a proxy for
circular velocity) and find that although minimized (particularly in the inner
parts of the halo where the difference in radial concentration is largest),
the bias still exists.  

The disruption and tidal stripping of satellites has recently been addressed
by a number of studies. Using controlled dark matter only simulations and
modeling the baryons in an idealized manner, 
\cite{2009arXiv0907.3482D} argue that the dominant mechanism in destroying
substructures in the inner regions of a galactic halo is tidal
shocking. However, their results fail to incorporate the full adiabatic effect
of baryons on the potential well of substructures. Being purely dark matter,
their substructures are more fragile and thus less able to hold onto their
material. %YH
Our results are in agreement with Romano-Diaz et al 2008, who found that the
number of substructures within $\sim 0.1~r_{vir}$ in the SPH case exceeds
the number in the pure DM case by almost a factor of two. 
%YH

 Again, in the context of pure dark matter simulations,
\cite{2009arXiv0907.0702W} looked at the disruption of satellites. They found
that the most disrupted satellites reside towards the centre of the halo, in
agreement with this work (see. Fig.~\ref{mlr}).

Although our study is focused on the constrained simulations of the local
group, it is difficult to compare our work with measurements of the radial 
concentration of the Milky Way's satellite population, because of the limiting
biases in the observations. The Sloan Digital Sky Survey, which increased the
number of satellites in the Milky Way's halo by a factor of two
\citep[see][and references therein]{2008ApJ...686..279K} covers a just a
single patch of sky. Additionally, this area may be biased as it points
towards Virgo: a direction that may have an over abundance of
substructures. However, as future sky surveys such as LSST and Pan-Starrs reveal to us a more complete satellite populations, we will obtain
a better understanding of the nature of the spatial distribution of the
companions of the Milky Way. 

\section*{Acknowledgments}

NIL is supported by the Minerva Stiftung of the Max Planck Gesellschaft. AK is
supported by the MICINN through the Ramon y Cajal programme. 
GY would like to thank also MEC (Spain) for financial support
under project numbers FPA2006-01105 and AYA2006-15492-C03. The simulations were
performed and analyzed at  the Leibniz Rechenzentrum Munich (LRZ) and at the
Barcelona Supercomputing Center (BSC). We thank DEISA for  giving  us  access
to  computing resources in these centers  through the DECI projects  SIMU-LU
and SIMUGAL-LU. We acknowledge the LEA Astro-PF collaboration and the ASTROSIM
network of the European Science Foundation (Science Meeting 2387) for the
financial support of the workshop ``The local universe: from dwarf galaxies to
galaxy clusters'' held in Jablonna near Warsaw in June/July 2009, where part of
this work was done.

\bibliographystyle{plainnat}
\bibliography{./ref.bib}
\end{document}